\RequirePackage{fix-cm}
\documentclass[smallextended]{svjour3}       
\smartqed  
\usepackage{graphicx}
\newcommand{\be}{\begin{equation}}
\newcommand{\ee}{\end{equation}}
\newcommand{\ben}{\begin{eqnarray}}
\newcommand{\een}{\end{eqnarray}}

\newcommand{\la}{{\lambda}}

\newcommand{\cO}{{\cal O}}

\newcommand{\cL}{{\cal L}}
\newcommand{\cH}{{\cal H}}
\newcommand{\cE}{{\cal E}}

\newcommand{\cJ}{{\cal J}}

\newcommand{\p}{\partial}
\newcommand{\na}{\nabla}

\newcommand{\ep}{\epsilon}

\newcommand{\tB}{{\tilde B}}
\newcommand{\tA}{{\tilde A}}

\begin{document}

\title{Mass formulae and staticity condition for {\it dark matter}  charged black holes}


\author{Marek Rogatko }


\institute{Institute of Physics  \\
Maria Curie-Sklodowska University\\
20-031 Lublin, pl.~Marii Curie-Sklodowskiej 1, Poland 
              \email{marek.rogatko@poczta.umcs.lublin.pl,~
rogat@kft.umcs.lublin.pl}          }

\date{Received: date / Accepted: date}

\maketitle
\begin{abstract}
The Arnowitt-Deser-Misner formalism is used to derive variations of mass, angular momentum and canonical energy for Einstein-Maxwell {\it dark matter}
gravity in which the auxiliary gauge field coupled via kinetic mixing term to the ordinary Maxwell one, which mimics properties of {\it hidden sector}.
Inspection of the initial data for the manifold with an interior boundary, having topology of $S^2$, enables us to find the generalised first law of black hole thermodynamics
in the aforementioned theory. It has been revealed that the stationary black hole solution being subject to the condition of 
encompassing a bifurcate Killing horizon with a bifurcation sphere, which is non-rotating, must be static and has
vanishing {\it magnetic} Maxwell and {\it dark matter} sector fields, on static slices of the spacetime under consideration.

\end{abstract}


\section{Introduction}
One of the long-standing puzzle
of the contemporary physics and astronomy is the {\it dark matter} problem,
ingredient which constitutes over 23 percent of the observable Universe, being an important
factor for the construction of cosmic web on which the ordinary matter accumulate \cite{mas07,die12}.

Nowadays {\it dark matter} sector interaction with the Standard model particles  \cite{reg15,ali15}
is widely elaborated. Several new types of fundamental particles were claimed to be the candidates
for the {\it hidden sector}. They are expected to interact with nuclei in detecting materials on Earth \cite{ber98}-\cite{cos18}.
On the other hand, there is a resurgence of implementing physics beyond the Standard Model
to explain {\it dark matter} interaction in non-gravitational sector \cite{mas15}-\cite{riz18}. 

The studies of gamma rays emission in dwarf galaxies, oscillation of the fine structure constant,
exploration of the possible low-energy mass {\it dark sector} \cite{ger15}-\cite{til15}, studies of {\it dark photon} production in the former observations of 
supernova 1987A event \cite{cha17}, Gaia-like astrometry observations \cite{guo19},
as well as, studies of dynamics of galaxy clusters collisions \cite{massey15a},
are expected to deliver new way of elaborating the old problem. Recently SENSEI device \cite{sen19} is dedicated for the direct-detection of {\it dark matter}
in eV to GeV energy range. It is expected that it provides constraints on {\it dark matter} electron 
scattering and on {\it dark photon} absorption of electrons.

However, the absence of evidences of the most popular particle candidates for the {\it dark matter} sector is affirmed, in some sense, as a
dilemma in the {\it hidden sector} researches. It has been proposed \cite{ber18} that one should carefully analyse upcoming astrophysical
phenomena, such like black hole and neutron star properties, gravitational wave signals to convey complimentary information connected with the
{\it dark matter} sector.

The problem in question, justifies also our researches connected with the influence of {\it dark matter} sector on black hole physics. In what follows
we shall try to find the mass formulae for black holes (especially effect of {\it hidden sector} on it), as well as, to prove the staticity theorem for black holes in 
{\it dark matter} gravity. As the staticity theorem imposes the additional conditions on the {\it dark matter} field, we suppose that the results will be of interests in further
inspection of {\it dark sector} properties.

In our work we shall consider
the model of {\it dark matter} sector in which the additional $U(1)$-gauge field is coupled to the ordinary Maxwell one. This the so-called {\it dark photon} model.
The action describing Einstein-Maxwell {\it dark matter} gravity \cite{vac91,ach00}  is provided by
\be
S = \int d^4x \sqrt{-g}~\bigg( R - F_{\mu\nu}~F^{\mu\nu} - B_{\mu\nu}~B^{\mu\nu} - \alpha~F_{\mu\nu}B^{\mu\nu} \bigg),
\label{gra}
\ee
where $R$ is the spacetime scalar curvature, $g_{\mu \nu}$ is four-dimensional metric tensor, $\sqrt{-g}$ stands for the square root of its determinant,  while
$F_{\mu \nu} = 2 \na_{[\mu }\tA_{\nu]}$ is the ordinary Maxwell field, $B_{\mu \nu} = 2 \na_{[\mu }\tB_{\nu]}$ is
 the additional $U(1)$-gauge field, mimicking the {\it {\it dark matter} } sector coupled to Maxwell one.  The auxiliary $U(1)$-gauge field is coupled
 to the Maxwell one by means of the so-called {\it kinetic mixing term} with a constant $\alpha$. The predicted value of $\alpha$, stemming from realistic
 string theory compactifications is between $10^{-2}$ to $10^{-16}$ \cite{abe04}-\cite{ban17}. The model described by the action (\ref{gra}) has its roots in
 string/M-theory, where the mixing portal arises quite naturally in open string theory. On the other hand, 
 in supersymmetric Type I, Type II A, Type II B models, both gauge states are supported by D-branes separated in extra dimensions \cite{ach16}.
 
In the action (\ref{gra}) the auxiliary gauge field is connected with some {\it hidden sector} \cite{ach16}.
In \cite{hol86} such model was used to describe the existence and subsequent integrating out of heavy bi-fundamental 
fields charged under the $U(1)$-gauge groups.
In general, such kind of terms emerge in the theories having in addition to some {\it visible} gauge group, the other one in
 the {\it hidden sector}. Such scenario is realised e.g., in compactified string or M-theory solutions generically possess
{|it hidden sectors}  which contain the gauge
fields and gauginos, due to the various group factors included in the gauge group symmetry of the {\it hidden sector}.

The {\it hidden sector} in the low-energy effective theory contains states which are uncharged under the the Standard 
Model gauge symmetry groups. They are charged under their own groups and interact with the {\it visible} ones via gravitational interaction. 
We can also think out other portals to our {\it visible sector} \cite{portal1,portal2}.
For the consistency and supersymmetry breaking \cite{abe08},
the realistic embeddings of the Standard Model in $E8 \times E8$ string theory, as well as, in type I, IIA, or IIB open string theory with branes, require the existence 
of the {\it hidden sectors}.

 As far as the possible experimental justification of the aforementioned theory is concerned, it happens that 
astrophysical observations of $511$ eV gamma rays
\cite{integral}, experiments detecting the electron positron excess in galaxies
\cite{liu15}-\cite{pamela}, as well as, possible explanation of muon anomalous magnetic moment \cite{muon}, 
strongly advocate the presented idea.
The {\it kinetic mixing term} between ordinary boson and relatively light one (the {\it dark} one) described as $U(1)$-gauge symmetry and connected with a {\it hidden sector},
may cause a low energy parity violation \cite{dav12}, which envisages itself in the Higgs boson decays,
and production of a relatively light vector boson with mass $m\ge 10GeV$ \cite{dav13}. \\

On the other hand, black holes are the other objects that still acquire attention in general relativity and its generalisations. The main motivation for our work is to
elaborate how {\it dark matter} sector influences black hole physics and characteristics like mass, angular momentum. We shall also consider how the {\it the hidden sector}
modifies the so-called staticity theorem which on its side constitutes the key starting point for the uniqueness theorem for black holes in the theory under consideration.

The staticity theorem was attributed to Lichnerowicz, who proved that a stationary fluid was everywhere locally static, i.e., its flow vector aligned with the Killing timelike one,
was static being the hypersurface orthogonal \cite{lic55}. 
Next extension was given in \cite{haw72}, where the vacuum non-rotating black holes were considered
in the sense that their null generator of the event horizon was aligned with Killing vector. 
The extension to electromagnetic case, subject to a certain inequality comprising
norm of the Killing vector field and electric field, was provided in \cite{car73}. The ADM formalism applied to the problem enables to get rid of the aforementioned inequality
\cite{sud92,sud93}. It was claimed that the solution of Einstein-Yang Mills is static if it had a vanishing Yang-Mills electric field on static slices of the considered spacetime.
The enhancement of the staticity theorem was provided in Einstein-Maxwell-axion-dilaton gravity being the low-energy limit of the heterotic string theory \cite{rog97,rog98}, and in higher dimensional Einstein-Maxwell gravity \cite{rog05}.
Staticity theorem plays an important role from the point of view of the uniqueness theorem for black holes (classification of black holes). Namely
in non-rotating case staticity is required,  whereas uniqueness for stationary axisymmetric black hole solutions is established by demanding circularity.
Both staticity and circularity imposes additional requirements on fields in the underlying spacetimes.

As far as the validity of the staticity in black hole mathematical theory is concerned, one should begin with
a  key result in the theory, which constitutes the theorem conceived by Hawking, the so-called {\it strong rigidity theorem} \cite{haw72,haw73}. The power
of it lies in the fact that it binds two distinct independently defined notions. The notion of black hole event horizon and a Killing vector field.
This theorem is of the key importance in the classification of stationary black holes.
In Ref. \cite{chr97} a refinement of the proof
of {\it the strong rigidity theorem} avoiding the previous assumption concerning 
{\it maximal analytic extensions} which were not unique, was elaborated. Further generalisation of {\it the rigidity theorem}for a spacetime with a stationary event horizon or a compact Cauchy horizon was given in \cite{fri99}.

It turns out that the uniqueness theorems for black holes are based on stronger assumptions than the {\it strong rigidity theorem}. As was mentioned above in the non-rotating case one requires staticity whereas in the  rotating case the uniqueness theorem is established for circular spacetimes. However,
another problem concerning the staticity theorem is to establish the {\it stationarity in the strict sense} ({\it strictly stationarity}),
i.e., one wants to exclude ergoregions in the non-rotating case \cite{hai73}-\cite{hai75a}.

The main aim of the uniqueness theorems of black holes in general relativity
is to show that the static electrovac black hole spacetime are described by
Reissner-Nordstr\"om spacetime whereas the circular one is diffeomorphic to Kerr-Newman
spacetime. For the first time, in Ref.\cite{isr} the problem of classification
of non-singular black hole solutions was studied. Then, in \cite{mil73} - \cite{rob77} it was shown how to weaken the original assumption presented in \cite{isr}.
The most complete results were proposed in Refs.\cite{bun87,ru,ma1,he1,he93}.
The classification of static vacuum and electro-vacuum black hole solutions was finished 
in \cite{chr99a,chr99b}, while
by taking into account the near-horizon geometry, the last restriction that all degenerate components of the event horizon should have charges of the same signs,
was removed \cite{chr07}.

The uniqueness black hole theorem for stationary axisymmetric
spacetime turned out to be far more complicated task. It was elaborated in Refs.\cite{car73,rob75}, while the complete proof was found by
Mazur \cite{maz} and Bunting \cite{bun}.

On the other hand, the unification attempts such as M/string theory
caused the  
resurgence of works devoted to the mathematical aspects of the low-energy string theory black holes as well as higher dimensional ones. 
Tthe uniqueness of the black hole solutions in dilaton gravity was  proved in 
works \cite{mas93,mar01}, while the uniqueness of the static dilaton $U(1)^2$
black holes being the solution of $N = 4,~ d = 4$ supergravity
was provided in \cite{rog99}. The extension of the proof to the theory
to allow for the inclusion of 
$U(1)^{N}$ static dilaton black holes was established in \cite{rog02}.\\

The layout of the paper is as follows. In Sec. II, having in mind the attitude presented in \cite{sud92,sud93}, we provide some basic idea concerning the Arnowitt-Deser-Misner formulation of Einstein-Maxwell 
{\it dark matter} gravity. We find the Hamiltonian for the theory in question, by the Legendre transformation and give the evolution equations. In Sec. III
one treats the problem of canonical energy and momentum, where the asymptotically flat hypersurface possesses and does not have the inner boundary.
Sec. IV is devoted to the mass formula and staticity theorem for charged {\it dark matter} gravity. We assume that the black hole in question
will be stationary and has a bifurcate Killing horizon. The main aspect of these studies will be two-folded. Firstly we shall look for the mathematical aspect of the staticity theorem for
the black holes influenced by the auxiliary $U(1)$-gauge field, responsible for
the {\it hidden sector}. Secondly in the light of growing new astrophysical observations of black holes, we take into account the problem how {\it dark matter} will modify
the black hole characteristics like mass and angular momentum of these objects.

\section{Arnowitt-Deser-Misner formalism for {\it dark sector} gravity}
In this section we shall provide the basic idea for the $(3+1)$ formalism for the Einstein-Maxwell {\it dark matter} theory. 
The formalism in question
considers the four-dimensional geometry
to consist of a foliation of three-geometries. It enables to divide the metric into spatial
and temporal parts, i.e., the line element yields
\be
ds^{2} = - (N dt)^{2} + h_{ab} (dx^{a} + N^{a}dt)(dx^{b} + N^{b}dt),
\ee
where $N$ is the lapse function while $N^a$ stands for the shift vector for the constant time hypersurface in the underlying manifold.
The geometry of the
manifold can be described in terms of the intrinsic metric and the
extrinsic curvature of a three-dimensional hypersurface,
$N$ and $N^a$ relate the intrinsic
coordinate on one hypersurface to the intrinsic coordinates on a nearby
hypersurface. Spacetime is sliced into spacelike hypersurfaces with
each hypersurface labeled by a global time parameter \cite{adm}-\cite{wald}. 
The general covariance allows the great arbitrariness in the choice of aforementioned functions
$N^\mu =  (N, ~N^{a})$.

In the canonical formulation of the Einstein-Maxwell {\it dark matter} theory
the point in the phase space corresponds
to the specification of the fields
$(h_{ab}, ~\pi^{ab}, ~\tA_{i},~\tB_i,~E_i,~B_i)$ 
on a three-dimensional
$\Sigma$ manifold, where $h_{ab}$ denotes Riemannian metric on $\Sigma$,~
$\tA_i$ and $\tB_i$ are $U(1)$-gauge Maxwell and {\it dark sector} fields on 
the three-dimensional manifold, while $E_i$ and $B_i$ are respectively, Maxwell and {\it dark matter} electric fields in the evolved spacetime.
They constitute tensor densities quantities and imply the following relations:
\be
E_{k} = \sqrt{h}~F_{\mu k} ~n^{\mu},
\label{elm}
\ee
for the Maxwell {\it electric} field,
while the {\it dark matter} sector {\it electric} field $B_{i}$ is provided by
\be
B_{k} = \sqrt{h}~B_{\mu k} ~n^{\mu},
\label{eld}
\ee
where $n^{\mu}$ constitutes the unit normal timelike vector to the hypersurface $\Sigma$
in the underlying spacetime. 
 
 The Lagrangian of the Einstein-Maxwell {\it dark matter} theory is subject to the relation
 \ben
 \cL_{EM-{\it dark ~matter}}  = N \sqrt{h} \Big( {}^{(3)}R &+& K_{ab} K^{ab} - K^2  - F_{ab} F^{ab}  + \frac{2}{h} E_i E^i  \\ \nonumber
 &-& B_{ab} B^{ab}  + \frac{2}{h} B_i B^i
 - \alpha F_{ab} B^{ab}  + \frac{2 \alpha}{h} E_i B^i \Big),
 \een
where $K_{ab}$ denotes the extrinsic curvature.

The corresponding field momenta are achieved by variation of  the underlying Lagrangian
with respect to $\na_{0}h_{ab}$, ~$\na_{0}\pi_{ab}$,~ $\na_{0}\tA_{i}$,~$\na_{0}\tB_{i}$,
where $\na_{0}$ denotes the derivative with respect to time-coordinate. 

Performing the variations with respect to $\na_{0}h_{ab}$, one has that
the momentum $\pi^{ab}$ canonically conjugate to a Riemannian metric can be 
expressed by means of the extrinsic curvature $K_{ab}$ of $\Sigma $ 
hypersurface
\be
\pi^{ab} = \sqrt{h} \left ( K^{ab} - h^{ab}K \right ).
\label{}
\ee
On the other hand, the momenta canonically conjugate to the $U(1)$
gauge fields $\tA_i$ and $\tB_i$ are equal respectively to
\ben
\pi^{k}_{(F)} &=&\frac{\delta \cL_{EM-{\it dark ~matter}}}{\delta (\na_{0}\tA_{k})} = 4 E^k + 2 \alpha B^k,\\
\pi^{k}_{(B)} &=& \frac{\delta \cL_{EM-{\it dark ~matter}}}{\delta (\na_{0}\tB_{i})} = 4 B^k + 2 \alpha E^k.
\label{}
\een

The Hamiltonian for the Einstein-Maxwell {\it dark matter} theory will be defined by the Legendre transform. It is given by
\ben \label{}
\cH &=& \pi^{ij} ~\na_{0} h_{ij} +  \pi^{i}_{(F)} ~\na_{0} \tA_{i} + \pi^{i}_{(B)} ~\na_{0} \tB_{i} 
- \cL_{EM-{\it dark ~matter}} \\ \nonumber
&=& N^{\mu} ~C_{\mu} + \tA_{0} ~{\tilde \tA} + \tB_{0} ~{\tilde \tB} + \cH_{div},
\een
where $\cH_{div}$ is the total divergence and has the form as follows:
\be
\cH_{div} = D_{k} \Big( \tA_0~ \pi^k_{(F)} + \tB_0~\pi^k_{(B)} \Big)
+ 2 D_{a} \Big( {N_{b} \pi^{ab} \over \sqrt{h}} \Big),
\label{}
\ee
where $D_m$ denotes the covariant derivative with respect to the metric $h_{ab}$, on the hypersurface $\Sigma$.

Because of the fact that the time component of $U(1)$ gauge fields $\tA_0,~\tB_0$ do not posses associated kinetic terms, they can be regarded as
Lagrange multipliers. They will be subject to the generalised Gauss laws provided by
\be
{\tilde \tA} = - D_k \pi^k_{(F)} = 0, \qquad {\tilde \tB} = - D_k \pi^k_{(B)} = 0.
\ee
For the components $C^\mu$ one obtains the following expressions:
\ben \label{c0} \nonumber
C_0 &=& \sqrt{h} \Big( F_{ab} F^{ab} + B_{ab}B^{ab} + \alpha F_{ab} B^{ab} \Big)
+ \frac{2}{\sqrt{h}} \Big( E_k E^k + B_k B^k +  \alpha E_k B^k \Big) \\ 
&+& \sqrt{h} \Big( - {}^{(3)}R + \frac{1}{h} \Big(\pi_{ij} \pi^{ij} - \frac{1}{2} \pi^2\Big) \Big), \\ \label{ca}
C_a &=& 2 \Big( 2 B_{ak} B^k + \alpha B_{ak} E^k + 2 F_{ak} E^k + \alpha F_{ak} B^k \Big) - 2 \sqrt{h}~ D_a \Big( \frac{N_b \pi^{ab}}{\sqrt{h}} \Big).
\een
It turns out that the evolution equations can be formally derived using the {\it volume integral contribution} to the Hamiltonian in questions, denoted by $\cH_v$.
It has pure constraint form, given by
\be
\cH_v = \int_{\Sigma} d\Sigma~ N^\mu C_\mu.
\label{hv1}
\ee
To proceed further, let us implement the Hamiltonian principle and calculate the variations of the corresponding fields in the phase space of the of compact support. 
Finding arbitrary infinitesimal variations $(\delta h_{ab}, ~\delta \pi_{ab},~\delta \tA_i,~\delta \tB_i,~\delta E_i,~\delta B_i)$, after integration by parts,
we obtain the change of the Hamiltonian $\cH_v$ caused by the variations in question. Namely, it yields
\be
\delta \cH_v = \int_{\Sigma} d \Sigma  \Big(
P^{ab}~ \delta h_{ab} + Q^{ab} ~\delta \pi_{ab} + R^i ~\delta \tA_i + P^i ~\delta \tB_i + S^i ~\delta E_i + Q^i ~\delta B_i \Big),
\label{dhv}
\ee
where we have denoted
\ben \label{pab} \nonumber
P^{ab} &=& N \sqrt{h}~ a^{ab} + 2 N^m \Big( 2 B_{m}{}^a B^b + \alpha B_m{}^a E^b + 2 F_m{}^a E^b + \alpha F_m {}^a B^b \Big)\\ \nonumber
&+& \sqrt{h} \Big( h^{ab} D^m D_m N - D^a D^b N \Big)
- \cL_{N^i} \pi^{ab}, \\  \label{qab}
Q_{ab} &=& \frac{N}{\sqrt{h}} \Big( 2 \pi_{ab} - \pi_k{}^k~h_{ab} \Big) + \cL_{N^i} h_{ab},\\ \label{ri}
R^i &=& -4 \sqrt{h} D_a \Big[ N \Big( F^{ai} + \frac{\alpha}{2} B^{ai}\Big) \Big] - 4 \cL_{N^i} \Big( E^i + \frac{\alpha}{2} B^i \Big),\\ \label{pi}
P^i &=& -4 \sqrt{h} D_a \Big[ N \Big( B^{ai} + \frac{\alpha}{2} F^{ai}\Big) \Big] - 4 \cL_{N^i} \Big( B^i + \frac{\alpha}{2} E^i \Big),\\ \label{sk}
S_k &=& \frac{4 N}{\sqrt{h}} \Big( E_k + \frac{\alpha}{2} B_k \Big) + 4 \Big( \cL_{N^i} \tA_k + \frac{\alpha}{2} \cL_{N^i} \tB_k \Big)
+ 2 D_k \Big(N \tA_0 \Big) \\ \nonumber
&+& \alpha ~D_k \Big( N \tB_0 \Big), \\ \label{qk}
Q_k &=& \frac{4 N}{\sqrt{h}} \Big( B_k + \frac{\alpha}{2} E_k \Big) + 4 \Big( \cL_{N^i} \tB_k + \frac{\alpha}{2} \cL_{N^i} \tA_k \Big)
+ 2 D_k \Big(N \tB_0 \Big) \\ \nonumber
&+& \alpha ~D_k \Big( N \tA_0 \Big).
\een
For the quantity $a^{ab}$ one has
\ben \label{aab} \nonumber
a^{ab} &=& \frac{1}{h} \Big( 2 \pi^a{}_j \pi^{bj} - \pi_m{}^{m} \pi^{ab} \Big) - \frac{1}{2} h^{ab} \Big( \pi_{ij} \pi^{ij} - \frac{1}{2} \pi_m{}^m \pi_k{}^k \Big) 
+ {}^{(3)} R - \frac{1}{2} h^{ab} {}^{(3)} R
\\ \nonumber
&+& 2 \Big( F^{aj}F_j{}^b + \frac{1}{4} h^{ab} F_{ij}F^{ij} \Big) + 2 \Big( B^{aj}B_j{}^b + \frac{1}{4} h^{ab} B_{ij}B^{ij} \Big)\\ \nonumber
&+& 2 \alpha \Big( F^{aj}B_j{}^b + \frac{1}{4} h^{ab} F_{ij}B^{ij} \Big) \\ \nonumber
&+& \frac{2}{h} \Big(E^a E^b - \frac{1}{2}h^{ab} E_k E^k \Big) + \frac{2}{h} \Big(B^a B^b - \frac{1}{2}h^{ab} B_k B^k \Big)\\ 
&+&\frac{2\alpha}{h} \Big(E^a B^b - \frac{1}{2}h^{ab} E_k B^k \Big).
\een
$ \cL_{N^i}  E^k$, as well as $ \cL_{N^i}  B^k$ and $ \cL_{N^i}  \pi^{ab}$ depicts the Lie derivatives of tensor densities and are given by
\ben
\cL_{N^i}  E^k &=& \sqrt{h}~N^c D_c \Big(  \frac{E^k}{\sqrt{h}} \Big) - E^c D_c N^k + E^k D_c N^c,\\
\cL_{N^i}  B^k &=& \sqrt{h}~N^c D_c \Big(  \frac{B^k}{\sqrt{h}} \Big) - B^c D_c N^k + B^k D_c N^c,\\
\cL_{N^i}  \pi^{ab} &=& \sqrt{h}~N^c D_c \Big(  \frac{\pi^{ab}}{\sqrt{h}} \Big) - 2 \pi^{c ( a} D_c N^{b)} + \pi^{ab} D_c N^c.
\een
On the other hand, $\cL_{N^i} \tA_k,~\cL_{N^i} \tB_k$ and $\cL_{N^i} h_{ab}$ correspond to the ordinary Lie derivatives.

Having in mind the relation (\ref{dhv}), we arrive at the following forms of the evolution equations for the Einstein-Maxwell {\it dark matter} system:
\ben
{\dot{ h}}_{ab} &=& \frac{\delta \cH_v}{\delta \pi^{ab} }= Q_{ab},\qquad
{\dot{\pi}}_{ab} = - \frac{\delta \cH_v}{\delta h^{ab}}= - P_{ab}, \\ \label{aaa}
{\dot{E}}_k &=& - \frac{\delta \cH_v}{\delta \tA^k} = -R_k, \qquad
{\dot{\tA}}_k = \frac{\delta \cH_v}{\delta E^k } = S_k,\\ \label{bbb}
{\dot{B}}_k &=& - \frac{\delta \cH_v}{\delta \tB^k } = -P_k, \qquad
{ \dot{\tB}}_k = \frac{\delta \cH_v}{\delta B^k } = Q_k.
\een
As was mentioned in \cite{sud92,sud93} the quantities $N^\mu,~\tA_0,~\tB_0$ are viewed as non-dynamical variables which are not represented in the phase space of
Einstein-Maxwel {\it dark matter} theory. This fact enables us to choose them arbitrarily.
The choice of $N^\mu$ is caused by the evolution of the considered system one looks for. In what follows we shall be interested in the case when
$N^\mu$ will be connected with a time translation or rotation in the spacetime. On the other hand, $\tA_0$ and $\tB_0$ one restricts to the case of approaching 
angle-dependent quantities, when $r$-coordinates tends to infinity.

It worth mentioning that equation (\ref{dhv}) corresponds to the volume contribution to the Hamiltonian of the theory in question. In the case when $\Sigma$ is a manifold 
without boundary, the non-vanishing
surface terms should be taken into account, when the perturbations of $\cH_v$ satisfying the asymptotic boundary conditions at infinity, are examined.
They appear due to the integration by parts procedure. Just, if we want to put the Hamiltonian in question, into the form (\ref{hv1}), one
ought to get rid of the aforementioned surface terms by adding them with the opposite signs. Consequently, we arrive at
\ben
\cH = \cH_v &+& \int_{S_\infty} dS_k ~\Big[
N \Big( D^a h_a{}^k - D^k h_m{}^m \Big) + 2 \frac{N^b \pi^k{}_b}{\sqrt{h}} \\ \nonumber
&+& 4 N^m \Big( \tA_m + \frac{\alpha}{2} \tB_m \Big) E^k + 2 N \Big( \tA_0 + \frac{\alpha}{2} \tB_0\Big) E^k \\ \nonumber
&+& 4 N^m \Big( \tB_m + \frac{\alpha}{2} \tA_m \Big) B^k + 2 N \Big( \tB_0 + \frac{\alpha}{2} \tA_0\Big) B^k  \Big].
\label{ech}
\een
By virtue of the direct calculation it can be envisaged that for all asymptotically flat perturbations, as well as, for $N^\mu, ~\tA_0,~\tB_0$ fulfilling the auxiliary
conditions $r \rightarrow \infty$, one reaches the conclusion that variation $\delta \cH$ is equal to the right-hand side of the relation given by the equation (\ref{dhv}).

In what follows we shall examine the {\it asymptotically flat initial data}. Thus, one has that in an asymptotic region of hypersurface $\Sigma$, being diffeomorphic
to ${\bf R}^3 -B$, where $B$ is compact, the following fall-off conditions for the fields in the phase space will be fulfilled:
\ben
h_{ab} &\approx& \delta_{ab} + \cO \Big( \frac{1}{r} \Big), \\
\pi_{ab} &\approx& \cO \Big( \frac{1}{r^2} \Big), \\
\tA_i &\approx& \cO \Big( \frac{1}{r} \Big), \qquad \tB_i \approx \cO \Big( \frac{1}{r} \Big),\\
E_i &\approx& \cO \Big( \frac{1}{r^2} \Big), \qquad B_i \approx \cO \Big( \frac{1}{r^2} \Big).
\een
Moreover, the standard asymptotic behavior of the lapse function, $N \approx 1 + \cO(1/r)$, and shift function
$N^a \approx \cO(1/r)$, will be supposed.

\section{Canonical energy, canonical momentum and the first law of thermodynamics}
In this section the problem of canonical energy and momentum with respect to the first law of black hole thermodynamics in 
Einstein-Maxwell {\it dark matter} sector gravity will be paid attention to.
We shall scrutinise the case when the hypersurface $\Sigma$ being asymptotically flat has no inner boundary, as well as, the context of $\Sigma$
admits interior boundary.

\subsection{$\Sigma$ does not possess inner boundary}

To commence with let us define the canonical 
energy on the constrained submanifold of the considered phase space, as the Hamiltonian function bounded with
the case when $N^\mu$ is an asymptotical translation at infinity. In the aforementioned case we have that, $N \rightarrow 1,~N^a \rightarrow 0$.
Consequently, using the exact form of the Hamiltonian (\ref{ech}) with the asymptotical conditions for $N^\mu$, it yields
\be
\cE = \frac{1}{16 \pi} \cH (N \rightarrow 1,~N^a \rightarrow 0) = m + \cE^{(F-\alpha)} + \cE^{(B-\alpha)},
\ee
where $m$ is the ADM mass, defined as follows:
\be
m = \frac{1}{16 \pi} \int_{S^\infty} dS_k ~\Big( D^a h_a{}^k - D^k h_m{}^m \Big),
\ee
while the quantities $\cE^{(F-\alpha)} $ and $\cE^{(B-\alpha)}$ imply
\ben
\cE^{(F-\alpha)}  &=& \frac{1}{8 \pi} \int_{S^\infty} dS_k ~\Big( \tA_0 + \frac{\alpha}{2} \tB_0 \Big) E^k, \\
\cE^{(B-\alpha)} &=& \frac{1}{8 \pi} \int_{S^\infty} dS_k ~\Big( \tB_0 + \frac{\alpha}{2} \tA_0 \Big) B^k.
\een
One can also notice that $\cH_v$, as a pure constraint on the considered phase space, disappears.

The forms of $\cE^{(F-\alpha)} $ and $\cE^{(B-\alpha)}$ suggest that they are gauge dependent, because of the fact that both $\tA_0$ and $\tB_0$
may be chosen in an arbitrary way. However, as was stated in \cite{sud92}, if one takes into account a stationary solutions of Einstein-Maxwell {\it dark matter}
equations of motion, these components of gauge fields are determined by the auxiliary conditions ${\dot \tA}_k= 0,~{\dot \tB}_k = 0,~{\dot E}_k = 0,~{\dot B}_k = 0$,
for all the cases when $N^\mu$ is chosen to be the stationary Killing vector field.
For the choice of ${\tA}_0$, having in mind the equation (\ref{aaa}) that ${\dot \tA}_k= 0$, which in turn implies that $S_k=0$.
Contracting this equation with $\tA_0 + \alpha/2 \tB_0$, we get
\be
2 \Big( \tA_0 + \frac{\alpha}{2} \tB_0 \Big) D_k \Big(N (\tA_0 + \frac{\alpha}{2} \tB_0) \Big) + \cO \Big( \frac{1}{r^2} \Big) = 0.
\ee
On the other hand, having in mind that $N \rightarrow 1$, one can write
\be 
D_k \Big[ \Big( \tA_0 + \frac{\alpha}{2} \tB_0 \Big)^2\Big] +  \cO \Big( \frac{1}{r^2} \Big) = 0.
\ee
It leads to the relation $\p_k [\Big( \tA_0 + \frac{\alpha}{2} \tB_0 \Big)^2] = 0$, which shows that the magnitude of $\tA_0 + \frac{\alpha}{2} \tB_0 $
is asymptotically constant. By virtue of this we define
\be
V_{(F-B)} = \lim_{r \rightarrow \infty}\sqrt{\tA_0 \tA_0 + \alpha \tA_0 \tB_0 + \frac{\alpha^2}{4}\tB_0 \tB_0}.
\label{vfb}
\ee
The same procedure applied to the {\it dark matter} sector $U(1)$-gauge field. Namely for the choice of
${\tB}_0$, equation (\ref{bbb}) implies that ${\dot \tB}_k = 0$, which leads to the conclusion that $Q_k$ is equal to zero. Contracting
this relation with $\tB_0 + \alpha/2 \tA_0$, 
results in definition of the form as
\be
V_{(B-F)} =  \lim_{r \rightarrow \infty} \sqrt{\tB_0 \tB_0 + \alpha \tA_0 \tB_0 + \frac{\alpha^2}{4} \tA_0 \tA_0}.
\label{vbf}
\ee
To proceed further, let us contract ${\dot \tA}_k= 0$ with $\tA_0 + \frac{\alpha}{2} \tB_0 $ and take the divergence of the outcome. One gets that
\be
D_k D^k \chi_F^2 = \cO \Big( \frac{1}{r^4} \Big),
\ee
where we set
\be
\chi_F^2 = V_{(F-B)} ^2 + \frac{C_1}{r} + \cO \Big( \frac{1}{r} \Big).
\label{chif}
\ee
$C_1$ stands for a constant. Next, let us apply $D_r$ to the equation (\ref{chif}) for $\chi_F^2$. It implies
\be
2 N \Big( \tA_0 + \frac{\alpha}{2} \tB_0 \Big)~D_r \Big( N~\Big( \tA_0 + \frac{\alpha}{2} \tB_0 \Big) \Big) = - \frac{C_1}{r^2} + \cO \Big( \frac{1}{r^2} \Big).
\ee
Having in mind the exact form of the derivative $D_r$ from the relation ${\dot \tA}_k= 0$, we finally arrive at the following:
\be
\Big( \tA_0 + \frac{\alpha}{2} \tB_0 \Big) \Big( E_r + \frac{\alpha}{2} B_r \Big) = \frac{C_1}{r^2} + \cO \Big( \frac{1}{r^2} \Big).
\ee
Taking into account the equation (\ref{vfb}), we can define in the stationary case $\cE^{(F-\alpha)} = V_{(F-B)} Q_F$, where the charge \cite{chr87}
is of the form
\be
Q_F = \pm \frac{1}{4 \pi} \int_{S_\infty} dS \mid \Big(E_k + \frac{\alpha}{2} B_k \Big) r^k\mid,
\ee
where $r^k$ stands for the unit radial vector and the choice of $\pm$ signs depends on whether\\
\be
\lim_{r \rightarrow \infty} r^2 \Big(\tA_0 + \frac{\alpha}{2} \tB_0 \Big) \Big( E_m + \frac{\alpha}{2} B_m \Big) r^m = C_1
\ee
has the positive or negative value.

Consequently, the equation ${\dot \tB}_k = 0$ and the procedure described above, reveal that we arrive to the following relations:
\be
\chi_B^2 = V_{(B-F)} ^2 + \frac{C_2}{r} + \cO \Big( \frac{1}{r} \Big),
\label{chib}
\ee
where $C_2$ denotes a constant and the relation
\be
\Big( \tB_0 + \frac{\alpha}{2} \tA_0 \Big) \Big( B_r + \frac{\alpha}{2} E_r \Big) = \frac{C_2}{r^2} + \cO \Big( \frac{1}{r^2} \Big).
\ee
All these enable us to define the charge connected with the {\it dark matter} sector
\be
Q_B = \pm \frac{1}{4 \pi} \int_{S_\infty} dS \mid \Big(B_k + \frac{\alpha}{2} E_k \Big) r^k\mid,
\ee
with the choice of $\pm$ signs depending on whether
\be
\lim_{r \rightarrow \infty} r^2 \Big(\tB_0 + \frac{\alpha}{2} \tA_0 \Big) \Big( B_k + \frac{\alpha}{2} E_k \Big) r^k = C_2
\ee
is of the positive or negative value.

Just considering initial data corresponding to a stationary solution to Einstein-Maxwell {\it dark sector} gravity and supposing that $N^\mu$ is stationary Killing
vector field and choosing $\tA_0,~\tB_0$ in such way  that $\tA_k$ and $\tB_k$, as well as, $E_k$ and $B_k$ are time independent, we arrive at
the following {\it theorem}:\\
Let us suppose that $(h_{ab}, ~\pi^{ab}, ~\tA_{i},~\tB_i,~E_i,~B_i)$ constitute smooth data for a stationary, asymptotically flat solution for
Einstein-Maxwell {\it dark matter} system. Moreover, we assume that the hypersurface $\Sigma$ of the initial data set has no interior boundary and
$(\delta h_{ab}, ~\delta \pi_{ab},~\delta \tA_i,~\delta \tB_i,~\delta E_i,~\delta B_i)$ authorises smooth, asymptotically flat solution of linearised constraint 
relations for the underlying theory. Then one arrives at the following relation:
\be
\delta \cE = \delta m + \delta \cE^{(F-\alpha)} + \delta \cE^{(B-\alpha)} = 0,
\ee
where $\delta \cE^{(F-\alpha)} = V_{(F-B)} ~\delta Q_F$ and $\delta \cE^{(B-\alpha)} = V_{(B-F)} ~\delta Q_B$.

One can conclude that every stationary solution is an extremum of the ADM mass at fixed both {\it electric} Maxwell and {\it electric} {\it dark matter} sector charges.

\subsection{$\Sigma$ possesses the inner boundary}
At first, we pay attention to 
the canonical angular momentum $\cJ$, defined as a Hamiltonian function being subject to the condition that
$N^\mu$ is an asymptotical rotation. It this case $N \rightarrow 0 $, while $N^a \rightarrow \phi^a$, where $\phi^a$ is the appropriate Killing vector field bounded with the rotation.
On this account, it implies
\be
\cJ = -  \frac{1}{16 \pi} \int_{S^\infty} dS_k~\Big[ 4 \phi^m \Big( \tA_m + \frac{\alpha}{2} \tB_m \Big) E^k + 4 \phi^m \Big( \tB_m + \frac{\alpha}{2} \tA_m \Big) B^k
+ 2 \phi^m \pi^k{}_m \Big].
\label{j}
\ee
If we convert the above integral into the one calculated over $\Sigma$, one gets
\be
\cJ= -  \frac{1}{16 \pi} \int_{\Sigma} d \Sigma~\Big[
\cL_{\phi^i} h_{ab} \pi^{ab} + 4 B^k ~\cL_{\phi^i} \Big( \tB_k + \frac{\alpha}{2} \tA_k \Big) + 4 E^k~\cL_{\phi^i} \Big( \tA_k + \frac{\alpha}{2} \tB_k \Big) \Big] + \cJ_{\Sigma},
\label{js}
\ee
where we set $\cJ_{\Sigma}$ in the form as follows:
\be
\cJ_{\Sigma} = -  \frac{1}{16 \pi} \int dS_k \Big[ 2 \phi^m \pi^k{}_m + 4 \phi^m \Big( \tA_m + \frac{\alpha}{2} \tB_m \Big) E^k + 4 \phi^m \Big( \tB_m 
+ \frac{\alpha}{2} \tA_m \Big) B^k \Big].
\ee
In the relation (\ref{js}) the integral over $\Sigma$ vanishes 
due to the fact that Killing vector field $\phi^\mu$ is equal to its tangential projection \cite{sud92,sud93}. Thus if the hypersurface
in question has no inner boundary the canonical momentum for the axisymmetric case is equal to zero.

Let us suppose that the hypersurface $\Sigma$ has an asymptotic region and a smooth interior boundary $S$. Moreover
we set that $N \rightarrow 1, ~N^a \rightarrow \Omega \phi^a$, where $\phi^a$ is an axial Killing vector field, while $\Omega$ is a constant vale quantity.
We obtain the following relation:
\ben \label{sur} \nonumber
16 \pi \Big( \delta \cE - \Omega ~\delta \cJ \Big) &&= \int_\Sigma d \Sigma 
\Big( P^{ab}~ \delta h_{ab} + Q^{ab} ~\delta \pi_{ab} + R^i ~\delta \tA_i + P^i ~\delta \tB_i \\ 
&+& S^i ~\delta E_i + Q^i ~\delta B_i \Big) 
+ \delta (surface~ terms).
\een
In \cite{sud92}, it was revealed that one can take an asymptotically flat hypersurface $\Sigma$, intersecting the sphere $S$ of a stationary black hole, being
a bifurcation Killing horizon and taking into account $N^\mu$ equal to $\zeta^\mu =t^\mu + \Omega ~\phi^\mu$. Putting ${\dot{\tA}}_k = {\dot{\tB}}_k = 0,
~{\dot{E}}_k = {\dot{B}}_k = 0$, one finds that the integral of (\ref{sur}) disappears and only one surface term survives. This term is equal to 
$2 \pi \kappa \delta A$, where $\kappa$ stands for the surface gravity of the surface $S$, $A$ is the surface of $S$.
Now, by the fact that the aforementioned Killing vector field responsible for the rotation, is equal to its tangential projection, the defined above angular momentum 
is equivalent to the angular momentum of black hole.

We conclude that in the case of $S$ being smooth inner boundary of the hypersurface $\Sigma$, one reaches the statement as follows:\\
{\it Theorem}: \\
If
$(h_{ab}, ~\pi^{ab},~\tA_{i},~\tB_i,~E_i,~B_i)$ comprises the smooth data for a stationary, asymptotically flat solution for
a stationary black hole with bifurcation sphere lying on $\Sigma$, as well as,
$(\delta h_{ab}, ~\delta \pi_{ab},~\delta \tA_i,~\delta \tB_i,~\delta E_i,~\delta B_i)$ comprehends smooth, asymptotically flat solution of linearized constraint 
relations for the underlying theory, then
\be
\delta \cE = \delta m + V_{(F-B)}~\delta Q_F + V_{(B-F)}~\delta Q_B - \Omega ~\delta \cJ = \frac{1}{8 \pi } \kappa ~\delta A.
\label{1lth}
\ee
The relation (\ref{1lth}) encompasses the first law of charged {\it dark sector} black hole thermodynamics.

\section{Mass formula and staticity theorem for {\it dark matter} black holes}
In this section we shall elaborate the problem of mass for black holes and
staticity in Einstein-Maxwell {\it dark matter} gravity, in the spacetime representing
a stationary black hole with a bifurcate Killing horizon with a bifurcation sphere.

In our studies one will pay attention to the
stationary black hole with a bifurcate Killing horizon possesses a timelike
Killing vector field, approaching a time translation in the asymptotic region and Killing vector field $\zeta^\mu $ which is orthogonal to the bifurcation sphere and vanishes on it.
If $t^\mu$ fails to coincide with $\zeta^\mu $, then the spacetime has an additional, axial Killing vector field $\phi^\mu$, satisfying
\be
\zeta^\mu = t^\mu + \Omega ~\phi^\mu,
\ee
where $\Omega$ is the angular velocity of the black hole horizon, while the angular momentum of the black hole is defined respectively as
\be
I_{BH} = \frac{1}{16 \pi} \int dS^{\mu \nu} \ep_{\mu \nu \rho \sigma} \na^{\rho } \phi^{\sigma}.
\ee
The constraints of the aforementioned 
spacetime theory will be given on hypersurface $\Sigma$ being the maximal one, i.e. for which one has that $\pi_a{}^a = 0$. The possibility of implementing such
condition results from the theorem proved in \cite{chr94}, stating  that any stationary black hole solution with bifurcate Killing horizon admits an asymptotically
flat maximal hypersurface which is asymptotically orthogonal to $t^\mu$. Its boundary constitutes a bifurcate surface of the horizon.

To commence with, let us find for the hypersurface in question, the equation for the lapse function $N=- k^\mu n_\mu$, for any Killing vector $k_\mu$ and $n_\mu$
unit normal to the maximal hypersurface.
Having in mind the equation
for $C_0$, given by (\ref{c0}), it may be verified that 
\be
D_m D^m N = \rho_D ~N,
\label{nn}
\ee
where for the explicit form of $\rho_D$ one obtains
\be
\rho_D = \frac{\pi_{ab} \pi^{ab}}{h} + \frac{1}{2} \Big(
F_{ij} F^{ij} + B_{ij} B^{ij} + \alpha F_{ij} B^{ij} \Big)
+ \frac{1}{h} \Big( E_k E^k + B_k B^k \Big)  + \frac{\alpha}{h} E_k B^k.
\label{rod}
\ee
In the next step, we choose $N=\la = -t^\mu~n_\mu$. It satisfies the boundary conditions $\la \mid_S = 0$ and $\la \mid_\infty = 1$
By virtue of the maximal principle and integrating (\ref{nn}), one reach to the relation \cite{sud93}
\be
4 \pi m - \kappa A = \int_\Sigma d \Sigma ~\la~\rho_D.
\label{rd}
\ee
The above equation is the restriction for the mass of stationary black hole with bifurcate horizon in charged {\it dark sector} gravity. 

From the explicit
form of $\rho_D$ given by equation (\ref{rod}), one can see that there are additional terms (comparing to Einstein-Maxwell gravity) connected with {\it dark sector},
in the part encompasses strength and electric fields of the gauge fields in question. 
Let us analyze the sign of $\rho_D$. Having in mind a general inequality inequality of the form $(F_{ij} - B_{ij})(F^{ij} - B^{ij}) \ge 0$, one can deduce that
$F_{ij} F^{ij} + B_{ij} B^{ij} \ge \alpha F_{ij} B^{ij}$. On the other hand, for the third and fourth term on the right-hand side of (\ref{rod}), considering the inequality
$(E_i - B_i)(E^i - B^i) \ge 0$, one can show that $E_i E^i + B_i B^i \ge \alpha E_i B^i$. All these lead to the conclusion that $\rho_D$ is non-negative.

On the other hand, the generalisation of the above derivation, using the Raychaudhuri equation \cite{haw73}, reveals that in any spacetime being foliated by maximal hypersurfaces the lapse function fulfils the relation (\ref{nn}), with $\rho_D$
\be
\rho_D = \frac{1}{h} \pi_{ij}~\pi^{ij} + R_{\mu \nu}~n^\mu n^\nu.
\label{rod1}
\ee
Further it implies that for a stationary black hole with a bifurcate horizon, equation (\ref{nn}) is satisfied and $\rho_D$ is given by (\ref{rod1}). Moreover $\rho_D$
will be non-negative provided that the matter fields satisfy the strong energy condition \cite{sud93}. It turned out that when $\rho_D$ is non-negative
the maximum principle may be implemented and solutions of (\ref{nn}) are uniquely determined by their boundary and asymptotic values.

{\it Dark matter} sector intensively influences on black hole mass, not only as the auxiliary field can contribute (as it is expected from the outset) but
one can see that the {\it kinetic mixing term} plays also the significant role in the mass of the black objects.
Perhaps this fact may justify the existence of
supermassive black holes in the early Universe, when {\it dark matter} was in abundance and played the significant role in the formation of cosmic web on which the ordinary matter accumulated. As far as the observation of the objects in question is concerned,
recently it has been announced \cite{mat19}
that 83 new supermassive black holes have been discovered, at the distances when the Universe was only 5 percent of its current age.
The formation of such massive objects soon after the Big Bang is still a tentative question and a real challenge for contemporary astrophysics and cosmology.

The similar results concerning the significant influence of the {\it hidden sector} on black hole mass have been revealed in \cite{rog18}, where
the physical version of the first law of black hole and first law of mechanics for the compact binary system is considered. On the other hand, the exact metric of static
spherically symmetric solution in Einstein-Maxwell charged {\it dark matter} gravity was obtained in \cite{kic19}.

Inspection of the equation (\ref{rd}) reveals that
because of the fact that $\la,~\rho_D$ are non-negative quantities, as in the case of Einstein Yang-Mills \cite{sud93}, Einstein-Maxwell-axion-dilaton gravity
 being the low-energy limit of the heterotic string theory \cite{rog98}, higher-dimensional Einstein-Maxwell gravity \cite{rog05}, we obtain the conclusion that mass of the black hole
 is greater or equal to the area of the bifurcate sphere multiplied by the surface gravity.
 The non-rotating case was treated in \cite{vis92}, where the inequality of this kind was revealed.

The other way of examining the mass formula is to exploit the relation derived in \cite{bar73}
\be
m = 2 \int_{\Sigma} d \Sigma \Big( T_{\mu \nu} - \frac{1}{2} g_{\mu \nu} T \Big) t^\mu n^\nu + \frac{\kappa~A}{4 \pi} + 2 \Omega~I_{BH}.
\ee
Implementing the exact form of the energy momentum tensor one obtains
\ben \label{rd1}
m &-& \frac{\kappa~A}{4 \pi} - 2 \Omega~I_{BH} = \frac{1}{4 \pi} \int_{\Sigma} d \Sigma \Big[
\frac{2 t^m}{\sqrt{h}} \Big(
F_{md} E^d + B_{md} B^d + \alpha~F_{md} B^d \Big) \\ \nonumber
&+& \frac{\la}{2} \Big( F_{ab} F^{ab} + B_{ab} B^{ab} + \alpha~F_{ab}B^{ab} \Big) + \frac{\la}{h} \Big(
E_iE^i + B_i B^i + \alpha E_i B^i \Big) \Big],
\een
where $\la = - t^\mu n_\mu$.

One should remark that the integral over $\Sigma$ in Eq.(\ref{js}) vanishes, due to the fact that axial Killing field $\phi^\mu$ is equal to its projection \cite{sud93}.
This fact implies that in our case we obtain that
\be
\cJ = \cJ_{\Sigma}.
\ee

Having in mind (\ref{j}) and (\ref{js}) and the relation \cite{sud93} $\na^\mu \phi^\nu = D^\mu \phi^\nu - 2 \phi_\rho n^{[\mu }K^{\nu ] \rho}$,
where $K_{\mu \nu}$ is the extrinsic curvature of the hypersurface $\Sigma$, one can draw a conclusion that $1/8 \pi \int_S dS_a \pi^{ab} \phi_b$
is equal to the black hole angular momentum $I_{BH}$. Computing the second term in (\ref{j}), we reveal that
\be
4 \pi \Omega ~\Big( I_{BH} - \cJ_\Sigma \Big) = - \int_S \frac{dS_k}{\sqrt{h}} \Big[
t^m \Big( \tA_m + \frac{\alpha}{2} \tB_m \Big) E^k + t^m \Big( \tB_m + \frac{\alpha}{2} \tA_m \Big) B^k \Big].
\ee
Consequently, for $N = \la$ we finally find that the following is fulfilled:
\ben \label{omm}
4 \pi \Big(  V_{(F-B)}Q_F &+& V_{(B-F)}Q_B \Big) + 4 \pi \Omega ~\Big( I_{BH} - \cJ_\Sigma \Big) \\ \nonumber
&=& \int_\Sigma \frac{d \Sigma}{\sqrt{h}}
\Big[ -\frac{\la}{4 \sqrt{h}} \Big( 4 E_k + 2 \alpha B_k \Big) E^k 
+ t^m \Big( F_{km} + \frac{\alpha}{2} B_{km} \Big) E^k \Big] \\ \nonumber
&+& \int_\Sigma \frac{d \Sigma}{\sqrt{h}} \Big[
-\frac{\la}{4 \sqrt{h}}
\Big( 4 B_k + 2 \alpha E_k \Big) B^k 
+ t^m \Big( B_{km} + \frac{\alpha}{2} F_{km} \Big) B^k \Big].
\een
On the other hand, the relations (\ref{rd}) and (\ref{omm}) and using the exact form of $\rho_D$, one receives the following:
\ben \label{omm1}
8 \pi \Big[ \Omega~ \cJ_\Sigma &-& \Big( V_{(F-B)} Q_F + V_{(B-F)} Q_B \Big) \Big] =\\ \nonumber
&=& \int_\Sigma d \Sigma ~\frac{\la}{h} \Big[
\pi_{ab} \pi^{ab} + 2 \Big( E_i + \frac{\alpha}{2} B_i \Big) E^i + 2 \Big( B_i + \frac{\alpha}{2} E_i \Big) B^i \Big].
\een
As in \cite{sud93}, we remark that the exterior region of {\it dark matter} sector black hole can be foliated by maximal hypersurfaces with a boundary $S$.
They are asymptotically orthogonal to the timelike Killing vector field $t^\mu$, when the strong energy condition is satisfied. Making use the equation (\ref{omm1})
to such kind of hypersurfaces, we have that $\pi^{ab} =0,$ and $E_k = B_k = 0$. Consequently, we get that $\la n^\mu$ is a Killing vector and $t^\mu = \la n^\mu$.
It provides the following {\it theorem}:\\
The stationary black hole solution of Einstein-Maxwell {\it dark matter} system with bifurcate Killing horizon satisfying the condition
\be
 \Omega~ \cJ_\Sigma - \Big( V_{(F-B)} Q_F + V_{(B-F)} Q_B \Big) =0,
 \label{sta}
 \ee
 is static and possesses vanishing {\it electric} Maxwell $E_k$ and {\it dark matter electric} $B_k$, fields on static slices through the underlying spacetime.

 Inspection of the equations (\ref{rd}),~(\ref{omm}) and the relation
\be
m + V_{(F-B)} Q_F + V_{(B-F)} Q_B - 2 \Omega \cJ_\Sigma = \frac{\kappa A}{4 \pi},
\ee
which follows from (\ref{1lth}) and the scaling properties \cite{sud92} of $m \rightarrow \beta m,~V_{(F-B)} \rightarrow V_{(F-B)},~V_{(B-F)} \rightarrow V_{(B-F)},
~\Omega \rightarrow 1/\beta \Omega,~\cJ_\Sigma \rightarrow \beta^2 \cJ_\Sigma,~\kappa \rightarrow 1/\beta \kappa,~ A \rightarrow \beta^2 A,~\tA_\mu \rightarrow
\beta \tA_\mu,~\tB_\mu \rightarrow \beta \tB_\mu$,
reveals that
\be
8 \pi ~\Omega~ \cJ_\Sigma = \int_\Sigma d \Sigma ~\la \Big( F_{ab}F^{ab} + B_{ab}B^{ab} + \alpha F_{ab} B^{ab} \Big) + \int_\Sigma d \Sigma ~
\frac{\la}{h} \pi_{ab} \pi^{ab},
\ee
and
\ben
8 \pi  \Big( V_{(F-B)}  Q_F &+& V_{(B-F)} Q_B \Big)  = \int_\Sigma d \Sigma \la \Big( F_{ab}F^{ab} + B_{ab}B^{ab} + \alpha F_{ab} B^{ab} \Big) \\ \nonumber
&-& 2 \int_\Sigma d \Sigma ~\frac{\la}{h} \Big[ \Big( E_i + \frac{\alpha}{2} B_i \Big) E^i + 2 \Big( B_i + \frac{\alpha}{2} E_i \Big) B^i \Big].
\een
Now we can formulate the staticity condition for Einstein-Maxwell {\it dark matter} gravity black holes.\\
{\it Theorem:}\\
A solution of Einstein-Maxwell {\it dark matter} gravity (with vanishing magnetic charges in {\it visible} and {\it hidden} sectors) describing a stationary black hole with a bifurcation Killing horizon with a bifurcation sphere,
for which $\Omega~ \cJ_\Sigma =0$ is static, being subject to the condition of vanishing Maxwell {\it magnetic} and {\it dark matter} magnetic fields
on a static slices of the considered spacetime manifold.

\section{Discussion and conclusions}
\label{sec3}
In our paper we have studied the $(3+1)$ formulation of the charged {\it dark matter} gravity, being generalisation of Einstein-Maxwell theory with additional $U(1)$-gauge
field coupled to the ordinary Maxwell one, by the so-called {\it kinetic mixing} term. In this so-called {\it dark photon } model, the auxiliary field represents the {\it hidden sector} and mimicked the properties of
{\it dark matter}.

The Arnowitt-Deser-Misner formalism enables one to find the variations of mass, angular momentum and canonical energy for the spacetime possessing the inner
boundary and without it. If the inner boundary is a spacetime of topology of a sphere $S^2$, we have found the generalised black hole first law of thermodynamics in which the terms connected with the {\it hidden sector} play the crucial role. 

Inspection of the equations (\ref{rod}), (\ref{rd}) and (\ref{rd1}), leads us to the conclusion that for the stationary black hole with a bifurcate Killing horizon,  
{\it dark matter} sector plays the significant role for value of its mass. The present results reveal that {\it hidden sector} not only contribute to the black hole mass
as the auxiliary field can do (as it is expected from the outset) but also via {\it kinetic mixing } term proportional to $F_{ab} B^{ab}$.

Maybe, due to the large abundance of {\it dark matter} in our Universe, this fact may 
contribute to the explanation of the existence of supermassive black holes in the early Universe, when {\it dark matter} played the significant role in construction of the scaffolding (cosmic web) on which the ordinary matter condensed. Black objects might gain their masses via collapse of ordinary matter and {\it hidden sector} fields coupled in non-trivial way to
Maxwell one.

Elaborating the maximal initial data for the aforementioned theory, we have found that stationary black hole being the solution of Einstein-Maxwell {\it dark matter} gravity,
with bifurcate Killing horizon, satisfying the relation
\be \nonumber
 \Omega~ \cJ_\Sigma - \Big( V_{(F-B)} Q_F + V_{(B-F)} Q_B \Big) =0,
 \ee
is static and possesses vanishing both Maxwell {\it electric} and {\it dark matter electric} fields on a static slices of the spacetime.
On the other hand, the black hole for which $\Omega~ \cJ_\Sigma =0$,
is static being subject to disappearing both Maxwell and {\it dark matter magnetic} fields on
static slices of the manifold in question.

These auxiliary conditions imposed on the adequate electric or magnetic components of the gauge fields, in the sense of definitions (\ref{elm}) and (\ref{eld}),
constitute the so-called staticity conditions for charged {\it dark matter} gravity enabling to achieve the static black hole solution in the theory.

The next aim will be to implement these restrictions to prove the uniqueness of static charged black holes with the {\it hidden  sector} field (find their classification), as well as, to consider the next step in construction of uniqueness theorem for stationary axisymmetric black hole solutions, circularity theorem. We hope to return to these subjects elsewhere.





\end{document}